# When Can Accessibility Help?: An Exploration of Accessibility Feature Recommendation on Mobile Devices

Jason Wu
jsonwu@cmu.edu
Carnegie Mellon University & Apple Inc.
Pittsburgh, PA, USA

Gabriel Reyes, Sam C. White, Xiaoyi Zhang, Jeffrey P. Bigham
{gareyes,samuel_white,xiaoyiz,jbigham}@apple.com
Apple Inc.
Cupertino, CA, USA

## ABSTRACT

Numerous accessibility features have been developed and included in consumer operating systems to provide people with a variety of disabilities additional ways to access computing devices. Unfortunately, many users, especially older adults who are more likely to experience ability changes, are not aware of these features or do not know which combination to use. In this paper, we first quantify this problem via a survey with 100 participants, demonstrating that very few people are aware of built-in accessibility features on their phones. These observations led us to investigate accessibility recommendation as a way to increase awareness and adoption. We developed four prototype recommenders that span different accessibility categories, which we used to collect insights from 20 older adults. Our work demonstrates the need to increase awareness of existing accessibility features on mobile devices, and shows that automated recommendation could help people find beneficial accessibility features.

## 1 INTRODUCTION

Over the past decades, numerous accessibility features have been developed for people with a wide variety of abilities to use computing devices. Screen readers present otherwise visual content audibly, zoom features enable people to see content better, switch controls allow people to navigate screen content with switches (triggers), features tune out noise and allow users to hear more of what matters, and reading support adds an auditory component to text while reading and/or writing. Although many features exist, it is unclear how or whether people find the accessibility features that they could benefit from.

There is reason to believe that many people do not know about these features, and thus may not discover these features when they need them. As an example, one study found that only 1 of 50 participants were aware of the zoom feature in their web browser [20]. Another study found that only 3 of 14 older adults were aware that many mobile devices now come with accessibility features [26]. Older adults present a case worth further investigating, as many of them may not consider themselves as having a disability, but may nevertheless benefit from accessibility features as they age [13].

In this paper, we explore methods for matching accessibility features for people who may not know they could benefit from them. A natural starting point is to consider whether recommender systems may help with discovering accessibility features. One challenge is obtaining the data needed to construct these models in a privacy-preserving way. Health-related, and especially disability-related, information is highly sensitive, and users may be unwilling to provide this data [16]. Moreover, common recommender systems work via collaborative filtering [41], which leverages the idea that users who like certain things will like similar things. This could work in the area of accessibility recommendation, *e.g.*, if a user has turned on the VoiceOver screen reader, we might infer that they might also want to turn on Audio Descriptions because other users often have that pair of features turned on together. Yet, one of the assumptions in this project is that many people will not know to turn on accessibility features at all. If a person has never turned on any accessibility feature, there is no usage data to even start the first recommendation (*i.e.*, the "cold start" problem in recommender system research [34]).

An alternative approach that we advance in this paper is to recommend accessibility features based on how a user is interacting with a device. For instance, if the user is holding the device closer (or farther) than we would expect, that might indicate that they are having trouble seeing it, and could thus benefit from a font size increase. Likewise, if users are unable to perform double-clicks fast enough for the gesture to be recognized, then we might suggest to them the feature that allows more time between clicks. This is not an entirely new idea, *e.g.*, if one presses the shift key on Microsoft Windows repeatedly (perhaps indicating they are having difficulty using it as a modifier key), then Windows will ask the user if they would like to turn on "Sticky Keys", an accessibility feature introduced in Apple System 6 (1988) that turns the modifier keys into toggles so key combinations can be pressed one key at a time [15]. It is also possible to detect stuttering in speech, which may eventually be connected to features to make speech recognition work better [36]. As we will present, many of the accessibility features available on today's smartphones can be activated using such mechanisms derived from behaviors detected based on device use.

The main assumption behind our approach is that users should use their device as they normally would in order to understand their usage patterns and receive recommendations for accessibility features. This differentiates our work from past work on, *e.g.*, digital games specifically designed to detect dyslexia [38, 40] or autism [44], and wizards that explicitly ask users to describe their accessibility needs. Therefore, our approach may not be able to recommend some accessibility features to a subset of users. For instance, our approach may not be able to recommend VoiceOver to a blind person, who has never heard of the VoiceOver screen reader feature. However, we might be able to recommend VoiceOver to a user who is using the zoom feature at a high zoom level and still holding the device closer than would be expected. Most accessibility features do not dramatically change smartphone functionalities and are amenable to our approach of recommending based on observed usage.

In this paper, we first present the results of a survey with 100 participants (including 25 people over the age of 50) demonstrating that very few people are aware of available accessibility features on their devices. We then select four common accessibility features, each selected from a main category of accessibility features (*e.g.*, vision, hearing, interaction and mobility), and develop prototype recommenders for them. We initialized our recommenders from a baseline study with 10 participants, and then used them to explore accessibility recommendation with 20 participants.
Our paper makes the following contributions:

- We demonstrate and quantify that awareness and knowledge of how to use accessibility features is low among smartphone users (one-fifth of users knew what "accessibility" means) and even lower among adults over the age of 50 (one-tenth), who are more likely to benefit from them.
- We show that many existing accessibility features are conducive to recommendation, and we provide recommendation strategies for detecting relevant usage behaviors. Using these strategies, we constructed four prototype recommenders spanning accessibility categories.
- We conducted a study with 20 participants to collect insights on the utility and preferences of accessibility feature recommendation.

## 2 MATCHING PEOPLE TO ACCESS TECHNOLOGIES

To inform our work, we first reviewed the existing space of matching people to access technology. Specifically, we examined: *(i)* matching by human experts, *(ii)* automated screening and detection, and *(iii)* automatic personalization approaches.

### 2.1 Matching by Human Experts
Traditionally, matching people to accessible technologies has been done by human experts or through recommendations of medical professionals. Assistive Technology (AT) specialists can be employed as consultants to provide guidance on making content (*e.g.*, educational curriculum) accessible [9]. Physicians or therapists may provide guidance on using assistive technology as part of rehabilitation therapy [10, 14], although this sort of support is not available to everyone who could benefit from accessibility features. Such matching is typically done in specialized environments and is costly in terms of time and money. Access technologies tend to have low adoption rates [24], perhaps because potential users do not have sufficient time to see how they would work into their daily lives. In contrast to this matching process, many people who could benefit from accessibility features on mobile devices they may already own may not know to seek external help or have access to it. Thus, our approach is focused on automatically detecting needs and proactively recommending potentially useful accessibility features that are already on their devices.

### 2.2 Automated Screening
Recent work has started to explore how to "detect" whether someone is likely to have a particular condition, which could be used to recommend that they consult an expert or even try out a particular assistive technology.

Many mobile health sensing efforts have focused on providing low-cost alternatives for monitoring chronic symptoms, such as asthma [35] and cystic fibrosis [30], or building mobile "screening" applications for detecting medical conditions [25, 44, 46]. Some screening applications reduce the dependence on a specialized procedure and involve completing a task, such as playing a game, or leveraging interaction data with specialized apps (*e.g.*, photo browser, lock screen) [39, 40, 42]. However, many of these approaches require active intervention on the user's part (*e.g.*, opening an app and performing a specialized screening procedure); as we later show, most people would not think to check their mobile device for accessibility affordances. In our case, we seek ways of passively detecting accessibility needs by monitoring natural interactions with unmodified applications. Downloading and using a screening app would indicate some knowledge of accessibility features.

### 2.3 Automatic Personalization
Once a usability or accessibility need is detected, it must be accommodated in the user interface. One approach is to generate a suitable interface using these parameters. SUPPLE automatically generated UIs that optimize applications for expected user interactions based on a set of device constraints and interaction traces [27]. SUPPLE was later extended to parameterize user ability in generating UIs for people with motor impairments [28]. To bootstrap their model, SUPPLE required explicit preference elicitation and ability modeling steps, which we seek to avoid. Accessibility features can be seen as mechanisms that allow user interfaces to be personalized (*i.e.*, both SUPPLE and the Zoom feature in iOS enable content to be displayed larger). Thus, recommending accessibility features is one way to determine the personalization that might benefit a particular individual.

Another way to think about when accessibility features might be useful is to consider a user's situation and dynamic needs [33].

For instance, using one's phone while walking may be seen as temporarily inducing a kind of visual and motor impairment [29]. However, even if impairments may be temporary (*i.e.,* situational impairments), it can be impractical to re-generate the user interface each time conditions change. A great deal of prior work has considered how to account for situationally induced impairments while doing various activities, such as walking [37] and driving [43]. Because situational impairments are by definition temporary in relation to situation, more work has considered how to detect when the situation warrants including some intervention (*e.g.,* detecting walking in order to make the font size bigger). We build from these ideas in recommending accessibility features by detecting accessibility needs and suggesting features that can present the UI more appropriately.

## 3 ACCESSIBILITY AWARENESS SURVEY

We conducted an online survey to understand people's awareness of accessibility features on their mobile phones. Specifically, we were interested in how people might react if they developed an accessibility need, and if they would know to use (or even check) their phones for features that could help. A challenge in investigating awareness of accessibility features is that questions directly asking about a feature, *e.g.,* "Did you know you can make the font size bigger on your phone?", may make respondents aware of the feature. To address this challenge, we used a staged reveal approach in which we first asked about hypothetical situations a feature could be useful in to see if participants mentioned accessibility features as a potential solution, and then later asked how they would go about configuring accessibility features to better support their needs.

In our survey, we asked participants about their smartphone usage and a series of questions aimed at assessing their awareness of accessibility features. We obtained 100 survey responses and collected demographic information (42M/58F, ages 19-83, mean age 42.4). Our sample included 25 adults above the age of 50, who are significantly more likely to develop a disability in the near term [13] and could benefit from accessibility features in the future. The survey was conducted through an online polling platform Pollfish [12] and targeted a general population in the United States. We asked respondents what type of smartphone they used most frequently — 42% used Apple iOS while 58% used Google Android. Most respondents (74%) were frequent smartphone users (*i.e.,* used their smartphone multiple times an hour), and 95% of respondents reported using their smartphones at least a couple of times per day. Furthermore, participants reported using a wide range of apps on their smartphone, with respondents reporting using Lifestyle (43%), Social Media (76%), Education (18%), Games/Entertainment (54%), Productivity (35%), Utility (56%), and News/Information (51%) apps. 62% of respondents reported wearing prescription glasses or contacts, and 18% reported that they were sometimes unable to hear clearly without the use of a hearing aid.

Our survey investigated the following research questions:

**RQ1** - Would users think to check their mobile device if they developed an accessibility need?
**RQ2** - Do users know how to configure their devices to better support accessibility needs?

To address these research questions, we included two types of questions in our survey: *(i)* hypothetical questions (**RQ1**), and *(ii)* feature-based questions (**RQ2**). Figure 1 shows the specific hypothetical scenarios and features used in these questions, which were chosen to span features from various categories (e.g., vision, hearing, etc.) and be plausible candidates for recommendation.

To answer **RQ1**, we first asked our participants how they would use their phone in hypothetical situations where they encountered certain types of impairments related to accessibility features. For example, we asked participants: "*Imagine your eyesight gets worse so you can't easily read what's on the phone screen, what would you do?*". This allowed us to infer awareness of these features, without revealing their existence in the question itself. In total, our survey contained 7 of these hypothetical questions (Figure 1), each corresponding to a different accessibility feature. Following the hypothetical questions, we presented feature-based questions (**RQ2**) that provided high-level solutions to some of the previously posed hypothetical situations but required respondents to demonstrate knowledge of feature usage. For example, we asked "*How can you make the content on the screen larger and easier to view?*". As a part of this set of questions, we also asked participants to describe what the function of the Accessibility menu was in their device's settings. In total, our survey contained 5 of these feature-based questions (Figure 1). Survey-takers were told to answer these questions using their existing knowledge of smartphone features (or indicate "I don't know") and were explicitly told not to search for answers online or check external resources. In summary, we purposefully staged our questions to gauge accessibility awareness without initially "giving away" the existence of accessibility features (*i.e.,* hypothetical questions); then, we "narrowed in" on specifically asking for device-based solutions (*i.e.,* feature-based questions).

Responses for all questions were coded by 3 researchers trained in HCI and qualitative methods. The responses were coded using the following categories.

- *Correct setting* (**C1**) - The response provided the correct accessibility setting (*e.g.,* Enable Larger Text in the Accessibility menu) or another feature that provided the same level of access to device content (*e.g.,* Change the system font size in the Display Settings). Solutions for any mobile operating system (*e.g.,* iOS, Android) were marked correct.
- *Other Smartphone-based solutions* (**C2**) - The response provided a solution on the smartphone but was either too vague (*e.g.,* Change the settings) or did not work in all cases the accessibility feature did (*e.g.,* Double-tap to zoom in). Responses in this category often indicated that the user was aware that their device was capable of making content more accessible, but did not know how to enable that functionality without additional help. One of the problems is that users need to know the name of the feature beforehand in order to search for it, and our goal is to proactively surface them to the user to remove this need.
- *Other* (**C3**) - The response provided did not demonstrate any knowledge of awareness or usage of mobile accessibility features, but, as we discuss later, can still constitute a valid course of action.

*Hypothetical Questions* (**RQ1**) - To analyze responses from our set of hypothetical questions, we wanted to see how many participants' responses mentioned using functionality already present on their smartphone (**C1**, **C2**). We coded the responses to the hypothetical questions with an inter-rater reliability (Fleiss' Kappa) of $\kappa = 0.68$. While there was some disagreement (between the **C1** and **C2** codes) on whether a response provided "the same level of access to device content" as the correct accessibility setting (*e.g.,* some responses mentioned using voice commands for questions targeted at touch accommodations), it was clear that users did not think to check their mobile devices for accessibility affordances (**RQ1**). Only 15.7% of participants would have attempted to look on their phone for a solution and only 12.1% identified the most effective setting (Figure 1).

We also more closely analyzed the **C3** category to better understand those responses. Unsurprisingly, a common response was that participants stated they would consult with a doctor or medical professional. For example, if their eyesight prevented them from easily reading on-screen content, they responded that they would go to an optometrist to get glasses. As previously mentioned, human experts can match access technology with a higher degree of certainty and effectiveness. However, for many reasons (*e.g.,* time/money requirements, doctor's lack of knowledge of a patient's device, non-medicalized accessibility need such as slightly degraded vision) this method may not lead to the widest adoption of accessibility features. Other participants proposed purchasing additional software or equipment to access content (*e.g.,* buying a loud Bluetooth speaker when unable to clearly hear content or buying a new phone). Finally, another common type of response included asking someone else to help perform the action or trying again slowly, but these would not be feasible in many situations.

*Feature-based Questions* (**RQ2**) - We performed a similar coding for our feature-based questions ($\kappa = 0.73$), but because participants were required to consider their smartphone as a part of their solution, we only focused on whether the response provided the Correct setting (**C1**). The responses from the feature-based questions indicate a similar conclusion, showing that on average only 10.3% of responses by participants mentioned the correct setting (or an alternative solution providing equivalent utility). Finally, to more directly answer **RQ2**, we asked participants to respond with the definition of "accessibility" in the context of their phone's settings, and only 18% responded correctly, suggesting that most people would not know how to access their devices' built-in accessibility capabilities.

In summary, the results from our awareness survey show that although users rely on their smartphones for a wide variety of tasks, they are generally unaware of the accessibility features available on their smartphone. While one might suggest that users seek out this information when they develop the need, they may not know to look, and continue to "get by" using their device (*e.g.,* squinting or bringing the phone closer to their eyes). Thus, we believe a smartphone that proactively recommends accessibility features would improve users' interactions with their devices.

## 4 RECOMMENDING ACCESSIBILITY

To chart the design space of how accessibility features could be recommended, we reviewed and categorized a set of features available on modern mobile platforms. For our exploration and building proof-of-concept prototypes, we scoped our effort to iOS 12 [1], which contains nearly 50 accessibility features (Figure 2). Many of its accessibility features (*e.g.,* font size adjustment, content magnification) are standard across other platforms, so our recommenders can be directly transferred. Our approaches to building recommenders (*i.e.,* strategies for feature recommendation) can also be applied more generally to features that we did not initially explore.

A few of these features either require special hardware (*e.g.,* hearing aids) to be connected in order to be used, or are meant to address accessibility needs that would likely prevent users from using the device without them (*e.g.,* VoiceOver), and so we would not be able to recommend these features. This is not intended to be an exhaustive or complete list, necessarily, but is rather the result of iterative process among the authors (accessibility and sensing experts) to identify promising directions for each feature. Generally, our approach to recommending accessibility involves *(i)* identifying an accessibility need (potentially corresponding to an available feature), *(ii)* developing a hypothesis about how it might manifest in device signals, (*e.g.,* sensor readings, system events), *(iii)* determining a detection strategy to decide when to surface a recommendation to the user, and *(iv)* empirically validating, to the extent possible, that the detection method works as intended with acceptable accuracy. In our work, we specifically focus on detecting potential accessibility needs from observed usage data. While detection constitutes a large part of recommendation, there are other aspects (*e.g.,* how to surface, strategies for increasing adoption) that we leave to future work.

We categorize accessibility detection into four approaches: statistical, near-miss, action sequences, and grouped detections. In this section, we give a brief description of each method along with an example use-case.

*4.0.1 Statistical.* Statistical detection involves identifying differences in users' behavior statistically over time. This approach is useful when one or more signals are known to be relevant, but it is unclear what specific bounds or behaviors to detect. For instance, users may not realize that they frequently hold the phone close to their face to read content on it or that they consistently listen to media at a high volume. This approach generally leverages statistical tests and outlier detection algorithms to compare an individual's usage patterns with a pre-defined range. This detection method draws from prior approaches, such as machine learning techniques for dyslexia detection [40] and ability detection [27, 28, 33]. Generally, such approaches have assumed labeled data for supervised machine learning or optimization algorithms; yet, we find that even when using simpler approaches with fewer data (*i.e.,* univariate statistical tests), we can successfully detect relevant behaviors.

> *Font Size Increase* – if a user tends to hold the phone closer (or farther) from their face than the common distance we expect, then they may benefit from a larger font size.

---

[1]https://www.apple.com/accessibility/iphone/

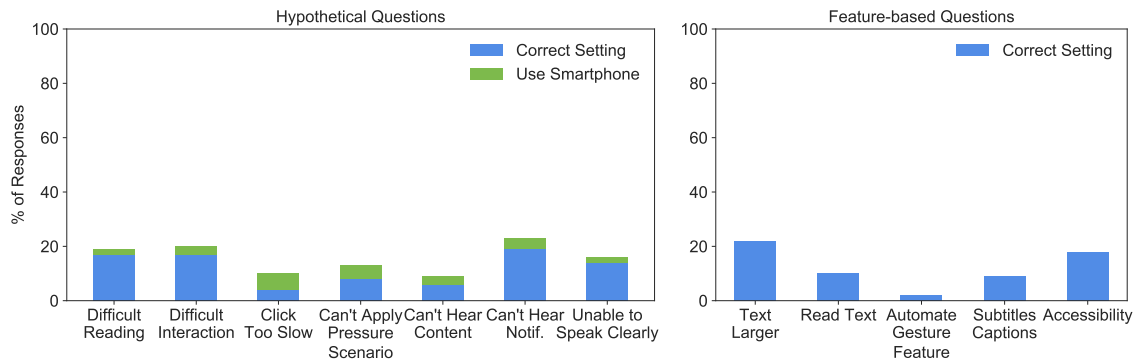

Figure 1: This figure shows a summary of responses to our online survey, which was composed of hypothetical (Left) and feature-based (Right) questions. For each of set of questions, we coded the responses to show people's knowledge of smartphone accessibility features. From our set of hypothetical questions (Left), we find that on average, 15.7% of participants would have attempted to find a solution on their smartphone and only 12.1% identified a setting that addressed the scenario. When asked which settings/features were needed to make certain content more accessible (Right), participants responded correctly 10.3% of the time, and only 18% of participants knew what "accessibility" meant in the context of their phone's settings.

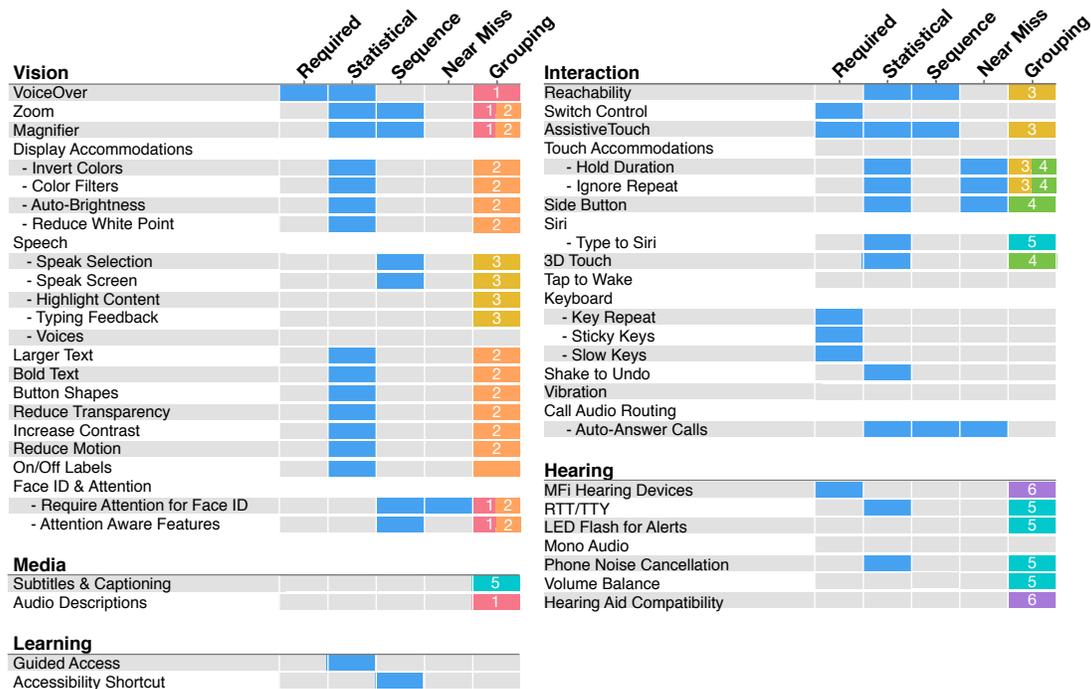

Figure 2: 48 accessibility features listed in the iOS Accessibility menu, labeled according to our detection strategies: *Required* are features that require hardware or whose users would be unlikely to use the device without them; *Statistical*, *Sequence*, *Near Miss*, and *Grouping* refers to the detection strategies. For every feature (row), a strategy (column) is highlighted if it is applicable. For features using the *Grouping* strategy, we also indicate which group they belong to (using a group number and color). For example, if the user has VoiceOver enabled, other features in that group (*e.g.,* Audio Descriptions, Require Attention for Face ID) can be recommended.

*4.0.2 Near-Miss.* Another type of detection strategy is to monitor features that rely on a pre-defined threshold or require multiple conditions to be reached before triggering. Often, the default threshold values may be difficult to reach for people with disabilities (*e.g.,* the default speed for double-clicking the home button may be too fast for people with motor impairments), so surfacing accessibility options that allow the adjustment of these values can significantly improve the user experience. Monitoring the threshold values for features that have them and logging "near-misses" can be used to trigger recommendations. Often, the accessibility feature itself gives a good clue as to what to look for, *i.e.,* if a double-click needs to happen with no more than 1 second latency between button presses, looking for examples of two button presses in sequence with a slightly longer gap between may be a good signal that the user could use more time.

> *Side Button Click Speed* – double and triple clicking the side button on iPhone invoke Apple Wallet and a shortcut to other accessibility features, respectively. Observing a user press the button two or three times, and fail to activate the feature (when they would have succeeded if the speed had been set to slower), may indicate that they could benefit from setting the required speed to slower.

*4.0.3 Action Sequences.* In some instances, a strong connection is known between a specific sequence of behaviors and a feature that might be useful. Action sequence detection is implemented by monitoring and recording system events (*e.g.,* app open events and UI interaction events). This method is informed by prior shortcut induction from user behavior [23], and programming by demonstration systems that have been used for accessibility purposes [18, 21, 22]. In most cases users may discover that some task is inaccessible and perform action sequences as "work-arounds" to achieve the same functionality, sub-optimally.

> *Magnifier* – the Magnifier in iOS allows users to use their phone's camera as a magnifier. We have informally observed people use an alternative way to access such functionality, which involves taking a picture of something (*e.g.*, a restaurant menu), opening the photo in the photo viewer, and then using "pinch to zoom". This sequence of actions could indicate that the user could benefit from the Magnification feature [8].

*4.0.4 Grouped.* Finally, grouped detection can be used to recommend new accessibility features based on the ones the user already has enabled. This is useful because related features do not always appear close to each other in the Settings menu, and some accessibility needs may not directly manifest themselves in signals detectable by other approaches. Grouping is a manual approach to replace recommendation algorithms, which do not have the necessary data to provide recommendations. Instead, we used simple conditional statements to recommend other grouped features. Unlike larger recommendation systems, the number of accessibility features to recommend is relatively small. A manually curated approach is manageable, although we envision grouped recommendations could potentially be data-driven if such data was available.

> *Type to Siri* – this feature allows users to type their queries to Siri. This may benefit deaf or hard-of-hearing users, and could be recommended when users turn on Hearing-related features.

Other examples are much more straightforward because they are already grouped together, *i.e.*, if you use one of the Vision-related features you might also benefit from other Vision-related features. For instance, neither Audio Descriptions nor Type to Siri is grouped with Vision or Hearing, respectively. In Section 5.4, we provide more details on the groupings implemented by our prototype system. Figure 2 shows more potential groupings between accessibility features, uniquely grouped by an identifying number and color.

## 5 PROTOTYPE RECOMMENDERS

To move towards concrete implementations of recommenders, we used several recommendation strategies and applied them to some of the features discussed in the accessibility awareness survey (Figure 1). We first performed a baseline data collection with an initial group of participants using a popular consumer smartphone (iPhone XS) to understand how different usage behaviors manifested themselves in sensor data. We describe the procedure for this data collection in the user studies section. Then, using our recommendation strategies, we built four accessibility feature recommendation prototypes. Each prototype targeted one or more accessibility features that could be detected using similar strategies. These prototypes are exemplars, which demonstrate how we might go about developing future recommenders.

### 5.1 Font Size Recommender

Our font-size recommender prototype automatically senses when users find it difficult to read content and recommend features that would adjust the content's size. As noted by previous research on text magnification, adjusting font size can enhance the readability and experience for many users, especially aging adults [20].

We calculated viewing distance using the front-facing camera and ARFaceAnchor objects returned by the ARFaceTrackingConfiguration [1] and used the *Statistical* detection strategy to surface recommendations to the user if they were found to hold the phone outside of an expected viewing distance range. We chose to define our expected viewing distance range empirically, based on our baseline data collection ($M_d = 0.36m, \sigma_d = 0.049m$). Our results align with previous work quantifying average font size and viewing distance for smartphone content [19]. We triggered a notification recommendation when the difference between the user's mean viewing distance and $M_d$ exceeded a threshold, which we conservatively set to two standard deviations.

### 5.2 Subtitles & Captions Recommender

Our "Subtitles and Captions Recommender" monitors device volume levels to recommend hearing accessibility features, similar to other features such as watchOS decibel meter, which does so for environmental noise [3]. We implemented a background daemon that continuously monitored 1) whether audio was currently playing, 2) the volume level, and 3) the output device. In the data collected from our baseline study, the average volume level was $M_v = 47.1\%, \sigma_v = 16.3\%$. Using the *Statistical* recommendation strategy, we surfaced a recommendation for the Subtitles & Captions feature when the user's listening volume was statistically greater (by a minimum of two standard deviations) than our baseline mean.

### 5.3 Side Button Click Speed Recommender

While touch interaction is the primary mode of interaction for most mobile devices, several important features require the use of physical buttons. These include adjusting output volume, locking/unlocking the device, and certain application-specific uses (*e.g.*, confirming an app installation) [6].

The default double-click speed on the side button can be difficult to trigger for many users with even slight motor impairments. Recognizing this, the time allowed between clicks can be changed (increased) via the accessibility menu [5]. Our prototype recommends this feature to users when it observes a "near-miss" failed attempt. To do this, the recommender monitors repeated button presses that occurred within the slowest possible double-click threshold. The recommendation is made if the input is too slow to trigger based on the current threshold, but would have done so using a slower setting.

### 5.4 Grouped Recommenders

Usage-based recommenders may be able to educate users about the existence of accessibility features, and then *Grouped* recommenders could help them expand and/or customize selected accessibility settings. Figure 2 shows a comprehensive grouping of accessibility features, while our grouped recommender prototype implements a subset of these using the iOS UIAccessibility API [2].

- AssistiveTouch → Side Button — If AssistiveTouch is enabled, the user might also benefit from the Side Button setting which can also make wider range of interactions accessible.
- Closed Captioning → Type to Siri — A Closed Captioning user may wish to interact with Siri using an alternative text-based modality.
- Bold Text → Larger Text — Similar to Bold Text, Larger Text increases the readability of on-screen text content by increasing the available font sizes.

## 6 USER STUDIES

We conducted two user studies: *(i)* a baseline data collection and *(ii)* a user study. As mentioned earlier, the baseline data collection with 10 participants aided in the design of our detection strategies and initializing our prototypes (*e.g.,* triggering thresholds). We then used our prototype recommenders with 20 participants to generate insight on the utility and preferences for accessibility feature recommendation. All user studies were conducted before the COVID-19 pandemic, so there was no additional risk for participants.

### 6.1 Baseline Data Collection

*6.1.1 Procedure.* We recruited 10 participants (7M/3F, ages 24-40, mean age 32) for our baseline data collection study. Eight of the participants wore glasses or contacts with corrective prescriptions, and no participants used hearing aids. After obtaining consent, participants were given an iPhone XS device that was preloaded with a background daemon that recorded a variety of signals.

At the start of every usage session, the researcher reset some of the device's settings (*e.g.,* output volume, system font size) to the lowest possible value. This was done to encourage the participant to set these values according to their own preference rather than using the ones chosen by the previous person, and the researcher informed participants that they were allowed to adjust these settings. To simulate everyday usage, we gave participants a pre-determined list of common smartphone tasks (Table 1) to complete during the study session. The order of the tasks was randomized for each participant. Participants were given 45 minutes to complete the list of tasks, and the data collection session was stopped if the tasks were finished early. Afterwards, participants were given a shortened version of the accessibility awareness survey that included questions about demographics, mobile phone usage, and the feature-based questions. Participants were compensated with a $10 gift card.

The data obtained helped inform the design of our prototypes and provided a dataset to empirically set our recommender algorithm parameters.

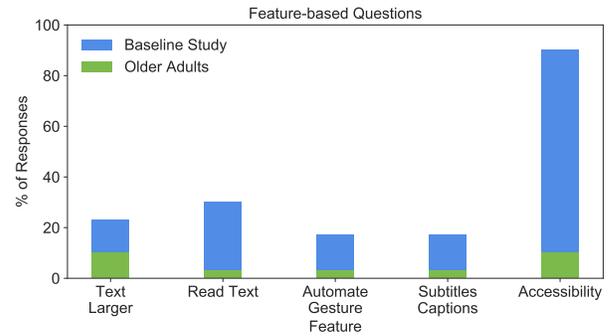

Figure 3: This figure shows the portion of responses from our in-person surveys (*i.e.,* Baseline Study and User Study) that provided the correct response to our feature-based questions. Awareness of features was much lower among adults over the age of 50, even though they were more likely to benefit from them.

### 6.2 User Study

To validate our approach and generate additional insights for accessibility recommendation, we conducted a study with two populations (n=20). All user studies were conducted before the COVID-19 pandemic, so there was no additional risk for participants. We were interested in answering the following research questions:

> **RQ1** - Were participants aware of the accessibility capabilities of their smartphones?
> **RQ2** - Did participants find the features recommended by our prototypes useful?

*6.2.1 Participant Age Range.* One of our motivations for this work was to surface recommendations for accessibility features to users who are likely to benefit from them (*e.g.,* older adults). In our recruitment of participants for this study, we settled on the age threshold of 50+, which we know is roughly the time when abilities start to really change [17, 45]. We acknowledge that this range is larger than most studies in HCI that study ageing, and likely encompasses several sub-groups (*e.g.,* older middle-aged, retirement-age adults, seniors) which have unique social norms, life experiences, and technology use. Our motivation is not bracket individuals into age groups or to create a solution specifically for one such group, but to identify broad segment of the population that can most benefit from accessibility recommendation.

*6.2.2 Procedure.* The first population (P1-P9) was recruited from a senior care residence and consisted of 9 participants (3M/6F, ages 79-97). We initially recruited 10 participants from our first population, but one later withdrew due to difficulty using the smartphone, and so we report on findings from the other 9 in that group. The second population (P10-P19) was recruited from a local participant pool with a minimum age requirement of 50 (5M/4F/1 Prefer not disclose, ages 50-79). Our participants had a diverse range of abilities, which allowed us to evaluate our system under circumstances experienced by a broad range of users. 78.9% of our participants wore glasses, and 36.8% participants used hearing aids. Most (78.9%) owned smartphones and reported varying frequency of usage — ranging from a couple of times per week or less (10.5%) to multiple times per hour (31.6%). When asked what kinds of apps participants used on their smartphones, some responded that they only used their phones for calling family members (5.3%), while others used Lifestyle (26.3%), Social Media (47.4%), Games/Entertainment (36.8%), Utility (57.9%), and News/Information (52.6%) apps. On average, participants reported that they had their smartphone for 4.3 years.

Table 1: This table describes the tasks participants performed during the data collection study.

| T # | Name | Description |
| --- | --- | --- |
| 1 | Video Questions | Watch a short 3-minute TED talk then fill out a quiz (24 questions) on the smartphone. |
| 2 | App Installation | Use the App Store to install 5 applications. After downloading and installing, take a screenshot of the main screen, then uninstall the app. |
| 3 | Internet Scavenger Hunt | Answer 9 trivia questions about GPS technology using provided external links and a search engine. Record answers in a note-taking app. |
| 4 | Siri Questions | Answer 7 questions using Siri and record answers in a note-taking app. |
| 5 | e-Reader Questions | Find 10 pieces of information in a book chapter and record answers in a note-taking app. |

For both populations, we followed a procedure similar to the one used for our baseline data collection (Table 1). Users were asked to complete a set of tasks during a usage session, and then filled out a survey afterwards. In this study, we shortened the usage session from 45 minutes to 30 minutes by removing two of the tasks (Internet Scavenger Hunt and the e-Reader questions) to make time for a brief interview afterwards about participants' views on the recommendations. In addition, due to some participants' lack of experience using smartphones and motor impairments, we reduced the complexity of some of these tasks (e.g., reducing number of questions on the video questionnaire). Although these procedural differences may impact the distribution of collected signals used for statistical detection (e.g., viewing distance), we only removed tasks if they were similar to others in the set (e.g., filling out an Internet Scavenger Hunt required typing text in the Notes app, as did the video questionnaire), and we did not observe a significant effect on our prototype's detection ability. For participants who preferred us to do so, we administered the post-study surveys verbally and recorded their answers for them in writing.

After administering the survey, researchers conducted a brief interview with participants structured around various accessibility features supported by our prototypes. We showed participants how to enable various accessibility features and demonstrated their effects on the user experience. We then asked participants to rate whether each feature was useful for them on a 7-point Likert scale (1: Strongly Disagree, 2: Disagree, 3: Somewhat Disagree, 4: Neutral, 5: Somewhat Agree, 6: Agree, 7: Strongly Agree). We concluded the interview by asking the participants if they thought they could benefit from accessibility features like these, and if so, how they would prefer the recommendations be surfaced. Participants were compensated $20 for their time.

### 6.3 Accessibility Awareness

Using the procedure for categorizing the questions from our online study, we coded the survey responses from both the Baseline Study and the Older Adults Study. The inter-rater agreement scores were $\kappa_{baseline} = 0.67$ and $\kappa_{senior} = 0.76$ for the responses from the Baseline Study and Older Adults Study, respectively. As the in-person surveys only included feature-based questions, we focused mainly on whether the responses fell into the Correct setting category (**C1**) or not. A comparison of the two populations can be seen in Figure 3. As seen in our analysis, most of our users were unaware of common accessibility features and how to use them. In answering **RQ1**, we found that there was very little awareness of accessibility features (or even what "accessibility" was) among older adults, even as they continue to engage with mobile technology (78.9% owned smartphones). Compared to the survey results from the baseline study, we found that awareness of accessibility features was much lower (10.5% of older adults knew what "accessibility features" were compared to 90% of baseline participants) among older adults, who more likely to benefit from them. Most of the participants in our Baseline Study were software engineers familiar with iOS and were more likely to know about accessibility features. Nevertheless, while 90% of participants generally knew about accessibility features, they were less familiar with specifics about how they could be used. From our Older Adults Study, we found relatively few participants knew about accessibility features (10.5%), even though they were more likely to benefit from them.

### 6.4 Utility of Accessibility Recommendation

Using the data collected from the participants' usage sessions, we ran our detection strategies post-hoc (i.e., participants did not interact with recommendations in real-time) to estimate the the utility of their recommendations. In total, our prototypes triggered 19 recommendations, and based on the participants' ratings of the features, they would have found 73.7% of those recommendations useful, 21.1% not useful, and 5.3% neutral. This suggests that our participants would likely benefit from accessibility feature recommendations (**RQ2**). Our goal was to gauge participants' perception feature recommendation and accessibility features overall, although additional work would need to be done to better understand the potential for these recommendations leading to adoption. Our conversations with participants support our initial analysis, and we aim to further strengthen this conclusion in future work.

Below, we further analyze each feature recommender in more detail and provide context for their performance. Because none of the participants had turned on any accessibility features, our Grouped recommender prototype is not applicable.

*6.4.1 Font Size/Zoom.* The Font Size/Zoom recommender was triggered by 3 of the participants (all 3 wore glasses), who, on average, gave those features a usefulness rating of 6.3/7. Interestingly, every participant in our study responded that they thought those features were useful when shown to them in our post-study interview. This may suggest that if such a feature recommender was deployed, it could be beneficial to decrease the triggering threshold or non-intrusively surface a recommendation for all users. Indeed, as part of the iOS new device setup process, a subset of display settings such as Display Zoom can be adjusted which affects the size of on-screen content. However, given that only two participants mentioned adjusting font size in their awareness survey responses, there is reason to believe that these features could be made more visible.

*6.4.2 Subtitles & Captions.* Similar to the Font Size recommender, the Subtitles & Captions recommender was also implemented by performing statistical detection on the user's audio volume. The mean audio levels of 6 of the participants (4 required hearing aids to hear clearly) exceeded our threshold, and those participants rated the Subtitles/Captioning features usefulness 4.8/7 on average. While the majority of triggered recommendations were found to be useful, we found that observed signals did not always align with participants' ratings of features. For example, P4 watched the video at maximum volume but rated the Subtitles & Captions feature as providing low usefulness. When asked about the rating, P4 responded that he disliked watching TV and movies with closed captioning because he found them distracting. Other instances of declined recommendations may be explained by preference or by prior work on the attitudes of older adults toward disability and aging [32]. We took from this that an additional element to consider when recommending a new feature is not only the expected utility of the feature but also the user's acceptance of it.

*6.4.3 Click Speed & Assistive Touch.* Compared to the other recommenders, the Click Speed & AssistiveTouch prototype was triggered the most often, in part due to the more relaxed *Near-Miss* detection scheme used. While in practice, such a system might surface a recommendation after a couple of near-misses, we set our prototype to trigger after the first instance due to the short duration of the study. In total, 10 users performed a double-click at speeds which would not have been detected using the *Default* timing but would have using slower settings. Of these, 70% found the associated accessibility features useful and 30% did not. Among users who triggered the recommender, the average usefulness rating was 4.7/7. However, for users that triggered our prototype's *Slowest* threshold, all of them (100%) found the features useful. While *Near-Miss* detection is appealing in part due to its simplicity and direct connection to an adjustable setting, successful deployment of recommenders for features such as Click Speed require more robust schemes that may combine certain aspects of statistical detection (*e.g.,* modeling the number of near-misses for the average user) and take into account additional context (*e.g.,* it is the first time the user is performing the double-click gesture).

## 6.5 Additional Observations

*6.5.1 Beyond Awareness.* An interesting observation was that one participant (P19) knew about accessibility features, but described them as *"for the visually/aurally impaired"*, which suggests that because he didn't identify as having a visual or hearing disability, he would not have thought of looking for useful features under the accessibility menu. We believe that a proactive recommendation system such as ours could help surface features that provide utility to a broad range of users. Indeed, two of our recommenders were triggered by the participant's usage session, and both features were marked as useful in the post-study interview. The only other participant (P11) who correctly described what "accessibility" referred to responded *"Can't remember ... But [I've] used [them]... maybe to increase text/magnify"*, indicating knowledge of only a small subset of features that could potentially be useful. Among participants recruited at the senior care residence (P1-P9), we found that knowledge of accessibility features was non-existent (*i.e.,* none of the participants from this group knew what "accessibility" meant in the context of computing), even though many of them had begun to adopt mobile technology (77.8% of them owned a smartphone, and on average, they owned their smartphone for 2.3 years). Similarly, we believe that our system could provide a lot of value for these technology adopters by making certain tasks easier to learn or perform.

*6.5.2 Recommendation Preferences.* After conducting the study, the researchers were frequently asked by participants to show them how to enable certain features on their personal devices. Even for participants who did not trigger any recommendations (P18), when shown certain features (*e.g.,* Font Size), they indicated that they thought the features would be beneficial: *"I have perfect eyesight but I still have a pair of readers that I use sometimes [to reduce strain] ... a bigger font size would also make things easier to read."* Almost all participants (89.5%) were open to receiving recommendations, with most preferring low frequency surfacing methods (*e.g.,* home screen, email, or a message) that did not interrupt their current task. On the other hand, one participant (P2) indicated that she valued the potential usefulness of features over the interruption cost: *"If there is something that could help me use [my device], I want to know about it."* Furthermore, not all participants wanted these features to be recommended to them (P6, P8). While P6 agreed that accessibility features were useful for interacting with her smartphone, she preferred not to have them automatically recommended to her, saying *"I might find it confusing"*. P8 offered another reason: *"Not necessarily... Once I learn [how to do something], I'll be set in my ways"*, stating that the novelty of interacting with the device through the recommended features might be off-putting. Similar to what we saw with declined recommendations, additional context such as user preference play an important role in user acceptance of accessibility features and new technology in general.

## 7 DISCUSSION AND FUTURE WORK

In this paper, we have introduced the idea of "recommending accessibility" as a fruitful area for research. Even as numerous useful accessibility features have started to be included in the smartphones that people own, our survey demonstrated that very few people know about them or know which of those features they could benefit from using. While on-device recommendation approaches are not the only useful strategy for building awareness, we believe that they are likely to be an important and necessary complement to existing advocacy and awareness approaches, especially as our devices, thankfully, contain more and more features intended to make them more accessible.

We have laid out a roadmap for recommending accessibility, explicitly outlining how accessibility features on iOS could map to a set of detection strategies and available signals that we believe largely characterize the space (Figure 2). We believe there is ample opportunity for future research in validating this across the many accessibility features on iOS and other platforms. Delivering a new recommender for a new feature will require substantial effort to validate, for instance, ensuring that the recommender works as intended and has an acceptable false positive rate in the real world. Ultimately, the success of a recommender approach seems dependent on a user deciding not only to turn on a feature but to adopt it; understanding adoption takes significant effort over time, especially given that each individual accessibility feature may only be expected to be useful to a relatively small percentage of people.

The prototypes introduced in this paper were designed as proof-of-concept implementations to demonstrate the potential of this direction and for use in our study with older adults. The accuracy and breadth of our recommendation prototypes can be further improved. We currently employ a simple statistical distance measure for unimodal data (*i.e.,* considering each signal independently). There is a great opportunity for future work to improve this aspect by developing more sophisticated methods of learning from usage data. Given a large enough sample, it may even be possible to learn retroactively from sensor streams collected from people who have turned on a feature.

Our approach to matching signals to detection strategies was largely manual. In some cases, we believe the mappings are fairly straightforward extensions of the accessibility feature specification, especially for the *Near-Miss* category that is explicitly defined this way. However, other categories, such as *Sequence* or *Statistical*, require generating hypotheses about differences exhibited by people who could benefit from the feature, collecting data to validate this hypothesis, and finally building a recommendation strategy based on that. Future work could investigate automated data-driven methods of identifying promising accessibility signals from usage patterns. In pursuing this, it is important to collect and analyze this data in a privacy-preserving manner. The *Statistical* signals rely on detecting differences from a collected baseline, and so developers of recommenders using this approach should be cognizant of where that baseline is collected from and be aware that some people may differ from the baseline for a reason other than needing the accessibility feature. Our framework does not provide a pattern for recommending all features that can be naively applied; future systems based on detecting and using accessibility signals should continue to rely on the intuition of designers, feedback from potential users, and iterative development and evaluation.

Another area to more thoroughly explore is how to surface accessibility recommendations. In our usage study, we presented users with recommendations after they completed their tasks, but recommendations can also be presented *in situ*. Notifications may be the most direct way of capturing attention and displaying information to users, but they can be disruptive or annoying [47]. Initiatives from Apple [7] and Google [11] have focused on limiting interruptions from notifications. The aforementioned "Sticky Keys" notification on Microsoft Windows has arguably been successful in getting people to know about the feature, but numerous web-based articles are devoted to turning off that notification (some fast-paced video games also often involve pressing modifier keys repeatedly and quickly). Less obtrusive ways of surfacing recommendations could include simply ranking the features higher in the accessibility menu, or including them in a non-intrusive but clearly visible place (*e.g.,* the lock screen) [4].

As a part of our exploration, we briefly explored different ways recommendations might be surfaced in a mobile operating system. While we expect more obtrusive notifications like pop-ups to be more likely attended to by users, they might also be more likely to disturb or annoy users[2]. Ultimately, there are many design decisions to be made, such as the wording and style used in the recommendation. In our prototypes, we specifically avoided relating the recommended feature to any underlying condition, ability, or cause. For example, we wrote recommendations like, *"Did you know you can adjust the font size?"* rather than *"It looks like you're having trouble seeing the screen."* We suspect that the preferred style and obtrusiveness of notifications may depend on the accuracy and timing of the recommendation, the likelihood that the user will follow the advice, and the realized benefit to them if they adopt the feature (*i.e.,* the classic expected utility problem in mixed-initiative interaction [31]). Ultimately, we believe the best strategy and frequency for surfacing recommendations will be feature and context dependent. We leave it to future work to empirically determine the best strategy for each feature and use case.

## 8 CONCLUSION

A large number of accessibility features have been developed for smartphone platforms. Our survey with 100 participants demonstrates that relatively few know about these features or how they might benefit from them. In this paper, we present our framework for recommending accessibility, outlining useful signals and detection methods using them to recommend accessibility features on the smartphone platform. We categorize 48 features on iOS in terms of how those features might be recommended to participants, and provide a number of example recommenders. We develop four prototypes, three of which we initialized in a baseline study with 10 participants. We then use our recommenders with another population of 20 participants to better understand their potential feasibility and utility. With so many great accessibility features being developed, we believe it is important to direct some of our research focus to recommending these accessibility features to those who might benefit from them. Our work provides a roadmap for researchers and developers to make progress in this important area.

## 9 ACKNOWLEDGEMENTS

We thank our participants and reviewers. We also thank Jeff Nichols for his feedback on earlier verions of this paper. This research was funded in part by a NSF GRFP Fellowship.

---

[2]Early in development, we also implemented a modal dialog box as an alert, but pilot studies guided us to less obtrusive designs.

# REFERENCES

[1] [n.d.]. Apple Developer ARFaceTrackingConfiguration. https://developer.apple.com/documentation/arkit/arfacetrackingconfiguration. Accessed: 2021-01-17.
[2] [n.d.]. Apple Developer Documentation UIAccessibility. https://developer.apple.com/documentation/uikit/uiaccessibility. Accessed: 2021-01-17.
[3] [n.d.]. Apple Newsroom watchOS 6 advances health and fitness capabilities for Apple Watch. https://nr.apple.com/d2i8t9c2n9. Accessed: 2021-01-17.
[4] [n.d.]. Apple Support About Siri Suggestions on iPhone. https://support.apple.com/guide/iphone/about-siri-suggestions-iph6f94af287/ios. Accessed: 2021-01-17.
[5] [n.d.]. Apple Support About the Accessibility Shortcut for iPhone, iPad, and iPod touch. https://support.apple.com/en-us/HT204390. Accessed: 2021-01-17.
[6] [n.d.]. Apple Support About the buttons and switches on your iPhone, iPad, or iPod touch. https://support.apple.com/en-us/HT203017. Accessed: 2021-01-17.
[7] [n.d.]. Apple Support Use notifications on your iPhone, iPad, and iPod touch. https://support.apple.com/en-us/HT201925. Accessed: 2021-01-17.
[8] [n.d.]. Apple Support Zoom in on the iPhone screen. https://support.apple.com/guide/iphone/zoom-in-on-the-screen-iph3e2e367e/ios. Accessed: 2021-01-17.
[9] [n.d.]. CforAT Accessibility Consultant. https://www.cforat.org/about/at_specialist. Accessed: 2021-01-17.
[10] [n.d.]. Craig Hospital Assistive Technology. https://craighospital.org/services/assistive-technology. Accessed: 2021-01-17.
[11] [n.d.]. Google Digital Wellbeing tools. https://wellbeing.google/tools/. Accessed: 2021-01-17.
[12] [n.d.]. Pollfish Real Consumer Insights. https://www.pollfish.com/. Accessed: 2021-01-17.
[13] [n.d.]. A Profile of Older Americans - 2017. https://acl.gov/sites/default/files/Aging%20and%20Disability%20in%20America/2017OlderAmericansProfile.pdf. Accessed: 2021-01-17.
[14] [n.d.]. Spaulding Rehab Assistive Technology. https://spauldingrehab.org/conditions-services/assistive-technology. Accessed: 2021-01-17.
[15] [n.d.]. Using Your Mac Classic. Accessed: 2021-01-17.
[16] [n.d.]. Web Axe Detecting Screen Readers - No. https://www.webaxe.org/detecting-screen-readers-no/. Accessed: 2021-01-17.
[17] [n.d.]. WHO Study on global AGEing and adult health (SAGE). https://www.who.int/healthinfo/sage/en/. Accessed: 2021-01-17.
[18] Michael Rivera Frank F. Xu Anhong Guo, Junhan Kong and Jeffrey P. Bigham. 2019. StateLens: A Reverse Engineering Solution for Making Existing Dynamic Touchscreens Accessible. In *Proceedings of the ACM Symposium on User Interface Software and Technology (UIST 2019)*.
[19] Yuliya Bababekova, Mark Rosenfield, Jennifer E Hue, and Rae R Huang. 2011. Font size and viewing distance of handheld smart phones. *Optometry and Vision Science* 88, 7 (2011), 795–797.
[20] Jeffrey P. Bigham. 2014. Making the Web Easier to See with Opportunistic Accessibility Improvement. In *Proceedings of the 27th Annual ACM Symposium on User Interface Software and Technology* (Honolulu, Hawaii, USA) *(UIST '14)*. ACM, New York, NY, USA, 117–122. https://doi.org/10.1145/2642918.2647357
[21] Jeffrey P Bigham, Jeremy T Brudvik, and Bernie Zhang. 2010. Accessibility by demonstration: enabling end users to guide developers to web accessibility solutions. In *Proceedings of the 12th international ACM SIGACCESS conference on Computers and accessibility*. ACM, 35–42.
[22] Jeffrey P Bigham, Tessa Lau, and Jeffrey Nichols. 2009. Trailblazer: enabling blind users to blaze trails through the web. In *Proceedings of the 14th international conference on Intelligent user interfaces*. ACM, 177–186.
[23] Allen Cypher. 1995. Eager: Programming repetitive tasks by example. In *Readings in human–computer interaction*. Elsevier, 804–810.
[24] Melissa Dawe. 2006. Desperately Seeking Simplicity: How Young Adults with Cognitive Disabilities and Their Families Adopt Assistive Technologies. In *Proceedings of the SIGCHI Conference on Human Factors in Computing Systems* (Montreal, Quebec, Canada) *(CHI '06)*. ACM, New York, NY, USA, 1143–1152. https://doi.org/10.1145/1124772.1124943
[25] Lilian de Greef, Mayank Goel, Min Joon Seo, Eric C. Larson, James W. Stout, James A. Taylor, and Shwetak N. Patel. 2014. Bilicam: Using Mobile Phones to Monitor Newborn Jaundice. In *Proceedings of the 2014 ACM International Joint Conference on Pervasive and Ubiquitous Computing* (Seattle, Washington) *(UbiComp '14)*. ACM, New York, NY, USA, 331–342. https://doi.org/10.1145/2632048.2632076
[26] Rachel L. Franz, Jacob O. Wobbrock, Yi Cheng, and Leah Findlater. 2019. Perception and Adoption of Mobile Accessibility Features by Older Adults Experiencing Ability Changes. In *Proceedings of the 2019 SIGACCESS Conference on Accessibility and Computing (ASSETS '19)*.
[27] Krzysztof Gajos and Daniel S. Weld. 2004. SUPPLE: Automatically Generating User Interfaces. In *Proceedings of the 9th International Conference on Intelligent User Interfaces* (Funchal, Madeira, Portugal) *(IUI '04)*. ACM, New York, NY, USA, 93–100. https://doi.org/10.1145/964442.964461

[28] Krzysztof Z. Gajos, Jacob O. Wobbrock, and Daniel S. Weld. 2008. Improving the Performance of Motor-impaired Users with Automatically-generated, Ability-based Interfaces. In *Proceedings of the SIGCHI Conference on Human Factors in Computing Systems* (Florence, Italy) *(CHI '08)*. ACM, New York, NY, USA, 1257–1266. https://doi.org/10.1145/1357054.1357250
[29] Mayank Goel, Leah Findlater, and Jacob Wobbrock. 2012. WalkType: using accelerometer data to accomodate situational impairments in mobile touch screen text entry. In *Proceedings of the SIGCHI Conference on Human Factors in Computing Systems*. ACM, 2687–2696.
[30] Gianluigi Grzincich, Rolando Gagliardini, Anna Bossi, Sergio Bella, Giuseppe Cimino, Natalia Cirilli, Laura Viviani, Eugenia Iacinti, and Serena Quattrucci. 2010. Evaluation of a home telemonitoring service for adult patients with cystic fibrosis: a pilot study. *Journal of telemedicine and telecare* 16, 7 (2010), 359–362.
[31] Eric Horvitz. 1999. Principles of mixed-initiative user interfaces. In *Proceedings of the SIGCHI conference on Human Factors in Computing Systems*. ACM, 159–166.
[32] Charlotte Humphrey, Katia Gilhome Herbst, and Shaista Faurqi. 1981. Some characteristics of the hearing-impaired elderly who do not present themselves for rehabilitation. *British Journal of Audiology* 15, 1 (1981), 25–30.
[33] Amy Hurst, Krzysztof Gajos, Leah Findlater, Jacob Wobbrock, Andrew Sears, and Shari Trewin. 2011. Dynamic Accessibility: Accommodating Differences in Ability and Situation. In *CHI '11 Extended Abstracts on Human Factors in Computing Systems* (Vancouver, BC, Canada) *(CHI EA '11)*. ACM, New York, NY, USA, 41–44. https://doi.org/10.1145/1979742.1979589
[34] Xuan Nhat Lam, Thuc Vu, Trong Duc Le, and Anh Duc Duong. 2008. Addressing cold-start problem in recommendation systems. In *Proceedings of the 2nd international conference on Ubiquitous information management and communication*. ACM, 208–211.
[35] Eric C Larson, Mayank Goel, Gaetano Boriello, Sonya Heltshe, Margaret Rosenfeld, and Shwetak N Patel. 2012. SpiroSmart: using a microphone to measure lung function on a mobile phone. In *Proceedings of the 2012 ACM Conference on ubiquitous computing*. ACM, 280–289.
[36] Colin Lea, Vikramjit Mitra, Aparna Joshi, Sachin Kajarekar, and Jeffrey P. Bigham. 2021. SEP-28K: A Dataset for Stuttering Event Detection from Podcasts with People Who Stutter. In *Proceedings of the IEEE International Conference on Acoustics, Speech and Signal Processing (ICASSP 2021)* (Virtual, McVirtualand).
[37] Terhi Mustonen, Maria Olkkonen, and Jukka Hakkinen. 2004. Examining mobile phone text legibility while walking. In *CHI'04 extended abstracts on Human factors in computing systems*. ACM, 1243–1246.
[38] Luz Rello, Ricardo Baeza-Yates, Abdullah Ali, Jeffrey P. Bigham, and Miquel Serra. 2020. Predicting risk of dyslexia with an online gamified test. *PLOS ONE* 15, 12 (12 2020), 1–15. https://doi.org/10.1371/journal.pone.0241687
[39] Luz Rello, Miguel Ballesteros, Abdullah Ali, Miquel Serra, D Alaron, and Jeffrey P Bigham. 2016. Dytective: Diagnosing risk of dyslexia with a game. In *Proceedings of Pervasive Health*, Vol. 16. http://www.luzrello.com/Publications_files/PerHealth2016-Dytective.pdf
[40] Luz Rello, Miguel Ballesteros, Abdullah Ali, Miquel Serra, Daniela Alarcón Sánchez, and Jeffrey P. Bigham. 2016. Dytective: Diagnosing Risk of Dyslexia with a Game. In *Proceedings of the 10th EAI International Conference on Pervasive Computing Technologies for Healthcare* (Cancun, Mexico) *(PervasiveHealth '16)*. ICST (Institute for Computer Sciences, Social-Informatics and Telecommunications Engineering), ICST, Brussels, Belgium, Belgium, 89–96. http://dl.acm.org/citation.cfm?id=3021319.3021333
[41] Badrul Munir Sarwar, George Karypis, Joseph A Konstan, John Riedl, et al. 2001. Item-based collaborative filtering recommendation algorithms. *Www* 1 (2001), 285–295.
[42] Feng Tian, Xiangmin Fan, Junjun Fan, Yicheng Zhu, Jing Gao, Dakuo Wang, Xiaojun Bi, and Hongan Wang. 2019. What Can Gestures Tell?: Detecting Motor Impairment in Early Parkinson's from Common Touch Gestural Interactions. In *Proceedings of the 2019 CHI Conference on Human Factors in Computing Systems* (Glasgow, Scotland Uk) *(CHI '19)*. ACM, New York, NY, USA, Article 83, 14 pages. https://doi.org/10.1145/3290605.3300313
[43] Omer Tsimhoni, Daniel Smith, and Paul Green. 2004. Address entry while driving: Speech recognition versus a touch-screen keyboard. *Human factors* 46, 4 (2004), 600–610.
[44] Tri Vu, Hoan Tran, Kun Woo Cho, Chen Song, Feng Lin, Chang Wen Chen, Michelle Hartley-McAndrew, Kathy Ralabate Doody, and Wenyao Xu. 2017. Effective and efficient visual stimuli design for quantitative autism screening: An exploratory study. In *2017 IEEE EMBS International Conference on Biomedical & Health Informatics (BHI)*. IEEE, 297–300.
[45] Morten Wahrendorf, Jan D Reinhardt, and Johannes Siegrist. 2013. Relationships of disability with age among adults aged 50 to 85: evidence from the United States, England and continental europe. *PloS one* 8, 8 (2013), e71893.
[46] Edward Jay Wang, William Li, Doug Hawkins, Terry Gernsheimer, Colette Norby-Slycord, and Shwetak N. Patel. 2016. HemaApp: Noninvasive Blood Screening of Hemoglobin Using Smartphone Cameras. In *Proceedings of the 2016 ACM International Joint Conference on Pervasive and Ubiquitous Computing* (Heidelberg, Germany) *(UbiComp '16)*. ACM, New York, NY, USA, 593–604. https://doi.org/10.1145/2971648.2971653

[47] Tilo Westermann, Sebastian Möller, and Ina Wechsung. 2015. Assessing the Relationship Between Technical Affinity, Stress and Notifications on Smartphones. In *Proceedings of the 17th International Conference on Human-Computer Interaction with Mobile Devices and Services Adjunct* (Copenhagen, Denmark) *(MobileHCI '15)*. ACM, New York, NY, USA, 652–659. https://doi.org/10.1145/2786567.2793684